# Quench Performance of the First Pre-series AUP Cryo-assembly

M. Baldini*[a], G. Chlachidze[a], G. Apollinari[a], J. Dimarco[a], S. Feher[a], V. Nikolic[a], D. Orris[a], R. Rabehl, S. Stoynev[a], T. Strauss[a], M. Tartaglia[a], A. Vouris[a]

*Abstract*—The High Luminosity upgrade of the Large Hadron Collider (HL-LHC) at CERN will include eight cryo-assemblies that are expected to be fabricated and delivered to CERN by the US HL-LHC Accelerator Upgrade Project (AUP) as part of the U.S. contributions to the HL-LHC. These cryostat assemblies are the quadrupole magnetic components of the HL-LHC Q1 and Q3 inner triplet optical elements in front of the two interaction points. Each cryo-assembly consists of two 4.2 m long $Nb_3Sn$ quadrupole magnets with aperture 150 mm and operating gradient 132.6 T/m. The first pre-series cryo-assembly has been fabricated and successfully tested at the horizontal test facility at Fermi National Accelerator Laboratory. In this manuscript we report the quench test results of the LQXFA/B-01 cryo-assembly. The primary objective of the horizontal test is full cryo-assembly qualification and validation of the performance requirements.

## I. INTRODUCTION

THE high luminosity upgrade of the Large Hadron Collider (HL-LHC) at CERN will include eight 10 m long cryo-assemblies - the Q1 and Q3 optical elements of the triplets at two LHC insertion regions. Each cold mass consists of two 4.2 m long (magnetic length) $Nb_3Sn$ high gradient quadrupole magnets with 150 mm aperture and nominal operating gradient of 132.6 T/m. Two MQXFA magnets are first tested in superfluid He at 1.9 K up to acceptance current of 16530 A (nominal current + 300 A) in accordance with functional requirements [1, 2] and, they are then installed in a stainless-steel helium vessel, including the end covers, to make the cold ass (LMQXFA). The LMQXFA cold mass, thermal shield and cryogenic piping inserted into the vacuum vessel makes the LQXFA/B cryo-assembly. The final configuration of the LQXFA/B cryo-assemblies (that are identical at this stage) suitable for LHC tunnel operation will be fabricated at CERN making the two distinct Q1 (LQXFA) and Q3 (LQXFB) optical elements. The fabrication and testing of the MQXFA magnets are the combined effort of three US Department of Energy laboratories: Brookhaven National Laboratory (BNL), Fermi National Accelerator Laboratory (FNAL), and Lawrence Berkeley National Laboratory (LBNL), which together comprise the Accelerator Upgrade Project (AUP). The AUP project is also responsible for the final design and fabrication of 10 cold masses (3 pre-series, 7 series production) including the design and construction of the cold mass fabrication tooling [3]. The fabrication of the cryo-assemblies using CERN provided cryostat kits is a US responsibility, together with the horizontal testing and shipping of the 10 cryo-assemblies to CERN [4]. The fabrication of the cold mass and the cryo-assembly and the horizontal tests are done at FNAL.

The horizontal test stand at Fermilab was previously used for testing the existing LHC inner triplet quadrupoles [5]. The cryogenic and mechanical subsystems of the stand were upgraded to meet the Q1/Q3 cryostat assembly design and test requirements. The old process controls, magnet protection and monitoring systems were also upgraded. Details of the horizontal test stand upgrades and commissioning are reported in [6]. Quench performance and magnetic measurements were carried out during the horizontal test. This manuscript focuses on the quench performance results of the first cryo-assembly LQXFA/B-01. Magnetic measurements are described in [7]. The acceptance criteria for the LQXFA/B-01 cryo-assembly [8, 9] were successfully met and the cryo-assembly design and fabrication processes were validated.

## II. CRYO-ASSEMBLY DESCRIPTION AND HORIZONTAL TEST FACILITY

The LQXFA/B-01 cryo-assembly [10] is made of LMQXFA01 [11] cold mass that contains two trained pre-series MQXFA magnets (MQXFA03 and MQXFA04 [12]). The magnet test parameters and cold test results of the two successfully trained magnets at the Vertical Test Facility at BNL are presented in [12]. The cryo-assembly cross section is shown in Fig. 1.

The LQXFA/B-01 cryo-assembly was cold tested at the horizontal test stand, also known as Stand 4 of the Fermilab magnet test facility [13]. A 30 kA power supply system [14] can power the magnets at the horizontal test stand using a water-cooled solid bus bar and water-cooled flexible buses. A picture of the LQXFA/B-01 cryo-assembly on the horizontal test stand is shown in Fig. 2. The test facility was refurbished to meet the new cryo-assembly design and test requirements [6].

This work was supported by the U.S. Department of Energy, Office of Science, Office of High Energy Physics, through the US HL-LHC Accelerator Upgrade Project. *(Corresponding author: Maria Baldini)*

M. Baldini*, G. Chlachidze, G. Apollinari, J. DiMarco, S. Feher, D. Orris, R. J. Rabehl, S. E. Stoynev are with the Applied Physics and Superconductivity Technology Directorate, Fermi National Accelerator Laboratory, Batavia, IL 60510, USA.

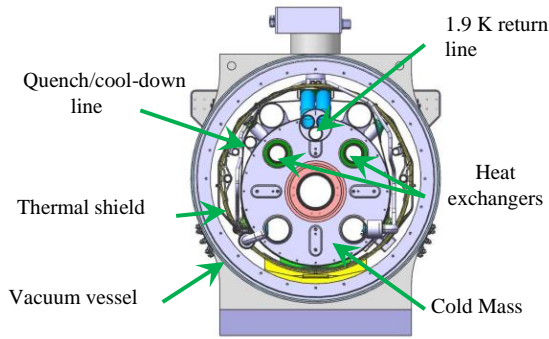

**Fig. 1.** Cross section of the LQXFA/B-01 cryo-assembly

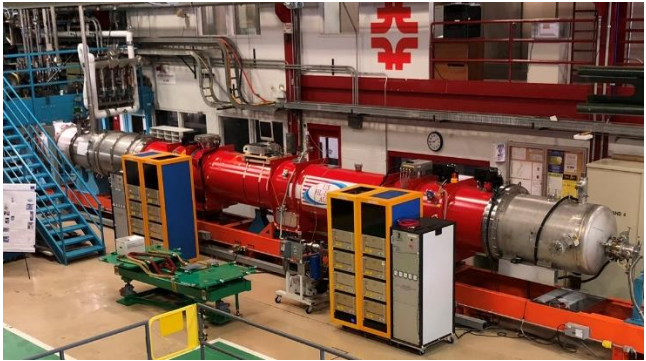

**Fig. 2.** LQXFA/B-01 cryo-assembly on the horizontal test stands. The HFU and CLIQ (Coupling-Loss Induced Quench) [15] units are also visible on the side.

The main requirements [8, 9] for testing of the LQXFA/B cryo-assembly are listed below:

- The nominal operating current is $I_{nom}$=16230 A, and the acceptance current is $I_{nom}$ + 300 A= 16530 A.
- Each MQXFA magnet shall be able to withstand a maximum temperature gradient of 50 K during a controlled warm-up or cool-down.
- The MQXFA magnets shall be capable of operating at any ramp rate within ±30 A/s and shall not quench while ramping down at 150 A/s from the nominal operating current.
- After a thermal cycle (TC) to room temperature, MQXFA magnets shall attain the nominal operating current with no more than 3 quenches.
- Splice resistance must be less than 1.0 nΩ at 1.9 K.
- LMQXFA cold mass shall be capable of continuous steady-state operation at nominal current in pressurized static superfluid helium (HeII) at 1.3 bar and at a temperature of 1.9 K. The requirement for LQXFA/B cryo-assemblies is to hold magnets at nominal current at 1.9 K for 300 minutes.

The main steps of the LQXFA/B cryo-assembly test program are comprised of warm and cold electrical checkouts, splice measurements, and quench training at 1.9 K.

### III. EXPERIMENTAL PARAMETERS, PROCEDURES, AND INSTRUMENTATIONS

The test procedure derives from an AUP magnet specifications document [1], which is based on the operational conditions of the magnets in the LHC triplets. Quench Detection system was based on detecting resistive voltages utilizing fixed voltage taps. Additional auxiliary (configurable) voltage taps were used for quench localization.

Two quench protection elements were employed during the horizontal test: Heater Firing Units (HFU) applied to the outer layer coil quench heaters (eight per magnet) [16-17] and utilizing Coupling-Loss Induced Quench (CLIQ) units [15]. The protection strategy has been designed to have the two magnets quenched via CLIQ and outer layer quench heaters, if a quench is detected in any of the MQXF magnets or buses. The HFU voltage and capacitance settings are fixed at 600 V and 7.05 mF respectively. The corresponding CLIQ settings are 50 mF and 600 V. Both protection elements are initiated at quench detection with no delay.

#### A. Warm and Cold Electrical Checkout

In Table I, the voltage withstand levels for the MQXF superconducting magnets at the horizontal test stand are reported for the warm and cold electrical checkouts [178]. The warm checkouts were performed at 300 K in air and the cold checkouts were first performed in liquid He at 4.5 K and then at 1.9 K after the second cooldown. One heater circuit in MQXFA03 magnet was found open during the cold mass assembly and consequently it was not used for the cold test. The heater circuit was isolated and removed from the protection.

TABLE I
VOLTAGE WITHSTAND LEVELS

| VOLTAGE LEVELS | Warm checkout (300 K) | Cold checkout (4.5 K/1.9 K) |
|---|---|---|
| COIL TO GROUND | 368 V | 1840 V |
| QUENCH HEATER TO COIL | 460 V | 2300 V |

The remaining quench heater circuits withstood the room temperature high voltage targets and passed the heater-to-coil and the coil-to-ground high voltage tests at room temperature.

High voltage withstand tests were also performed at 4.5 K. Another quench heater circuit of MQXFA03 failed the test at 2190 V. The heater circuit was isolated and not used. The rest of the heater circuits passed all the cold high voltage tests. In conclusion, six heater circuits in MQXFA03 and eight heater circuits in MQXFA04 were employed for quench protection during the LQXFA/B-01 horizontal test.

#### B. Cooldown

The controlled cooldown of the LQXFA/B-01 cryo-assembly started in March 2023.

The cooldown of LQXFA/B-01 was performed in three steps – down to liquid nitrogen (LN) temperature with a mixture of room temperature and 80 K helium gas, below LN temperatures with a mixture of 80 K He gas and liquid, and final cool down by liquid helium. The temperature was measured using two temperature sensors installed on the two extremities of the cold mass. In Fig. 3, the temperature of each end (lead and return end) of the LQXFA/B-01 cryo-assembly is reported together with the temperature differences between the ends. The temperature difference between the cryo-assembly ends was kept below 80 K. This constrain is sufficient to verify that a maximum temperature gradient of 50 K is present in each magnet. This result was confirmed calculating the temperature difference from the resistance values measured across each individual magnet during cooldowns. This temperature difference calculated using the resistance values is also reported in Fig.3

### IV. HORIZONTAL TEST RESULTS

After cooldown, the quench detection threshold settings for the voltage difference between magnets and half-magnets, together with the voltage signals from superconducting or copper leads were adjusted based on flux jump measurements at low currents.

#### A. Splice Measurements

The first measurements after the system verifications were the splice resistance measurements at 1.9 K with current ramped up to 5 kA. The NbTi-Nb$_3$Sn and NbTi-NbTi splice resistance values of the coil were already verified during the vertical test of each magnet and the measurements were not repeated during the horizontal test. All the splices made for the fabrication of the cryo-assembly were measured during this test, including splices, connecting MQXFA03 and MQXFA04 magnets.

Two additional splices were made to add extra length to each of the LQXFA/B-01 cryo-assembly leads for testing at Fermilab's horizontal test facility. It is important to note that this extra bus length is required only for testing at Fermilab and will be eliminated during the cryo-assembly installation at CERN.

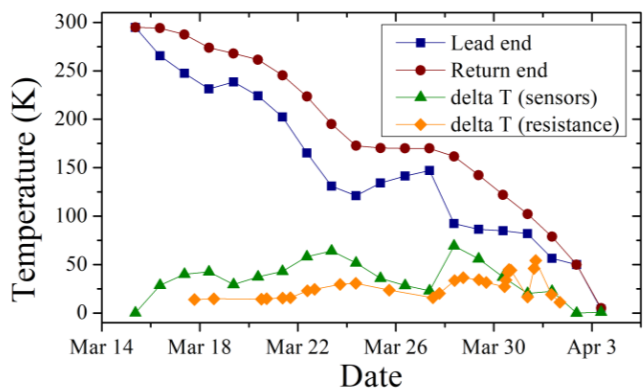

**Fig. 3.** Temperature measurements at the lead end (blue squares) and the return end (red dots) of the LQXFA/B-01 cryo-assembly. The temperature differences measured with sensors (green triangle) and calculated from resistance values (orange diamonds) are also reported.

The resistance values of the two-splice segments were determined to be 0.7±0.1 nΩ, and less than 0.3 nΩ for the single splice segments. All splice resistance values meet the acceptance requirements discussed before. The splice resistance data collected for one of the two-splice segments are shown in Fig. 4.

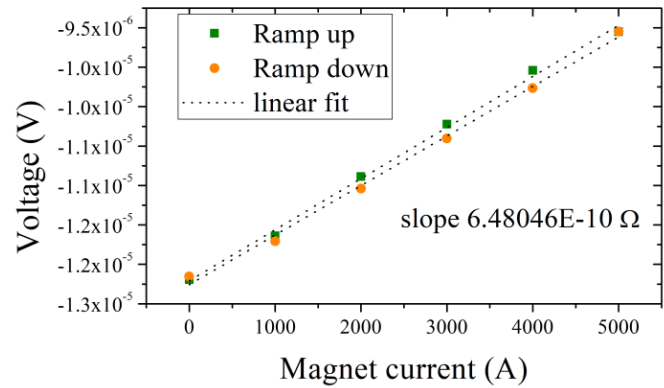

**Fig. 4.** Voltage measured across one of the two-splice segments as a function of magnet current. Instrumental offset is not removed on the plot.

#### B. First Cooldown: Quench Training Results

The quench training at 1.9 K was started with a current ramp rate of 20 A/s. Quench data acquisition sampling rate was 10 kHz. The first step of the quench training program was to ramp the cryo-assembly up to acceptance current and hold the current for 30 min. The second step was to ramp up to nominal current at 30 A/s and hold it for 5 min at the flattop. The quench performance during the first thermal cycle is shown in Fig. 5. The two MQXFA magnets in the LQXFA/B-01 cryo-assembly reached acceptance current without any spontaneous quenches. However, two quenches in the superconducting leads (outside of the cryo-assembly) were observed before reaching the acceptance current, at 15387 A and 16388 A, with 24.37 and 24.36 MIIts (MA$^2$s) accumulated, respectively. Those quenches in the leads were due to a low liquid helium (LHe) level. LHe level requirements were adjusted afterward. The 30 A/s ramp rate requirement was also successfully verified ramping the magnet up to the nominal current.

The hot spot temperature limit (250 K) during a quench determines the maximum MIIts budget. For MQXFA magnets, the corresponding quench integral ($\int I^2 dt$) value is about 32 MIIts (MA$^2$s) [12].

The next step of the test was to perform the five-hour holding current test. Keeping the current level at 16530 A an electrical failure (trip) of the power supply occurred after 33 min. A quench was observed in MQXFA04 during the subsequent ramp up. The quench took place at 16386 A (27.35 MIIts). The test to acceptance current (30 min holding at flat top) and to nominal current (5 min holding at flat top) was then successfully repeated. The five-hour holding current test failed again after 85 min because of the LHe level instability in the test stand. The ramp to acceptance current was successfully repeated afterwards and a thirty-minute holding current at 16230 A was successfully performed at 4.2 K.

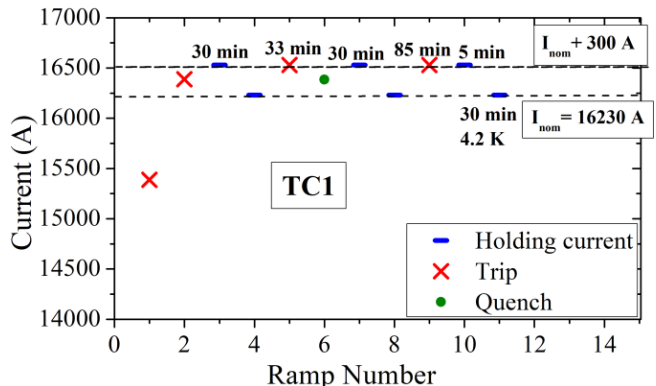

**Fig. 5.** Quench performance during the first thermal cycle (TC1).

Because of the instabilities in the cryogenic system the LQXFA/B-01 cryo-assembly was then warmed up. A mitigation measure was set into place for the second thermal cycle to avoid superconducting lead quenches due to sudden change of LHe level: if the liquid level was observed to drop below a specific threshold, a slow ramp down was initiated.

*C. Second Cooldown: Quench Training Results*

The second cooldown of the LQXFA/B-01 cryo-assembly was successfully performed, and high voltage tests were repeated at 1.9 K. All the quench detection settings were verified and adjusted as necessary. For example, the magnet voltage detection thresholds were set to higher values for high ramp rate studies. The cryo-assembly quench performance in the second thermal cycle is shown in Fig. 6. No spontaneous quench was observed, and several ramps up to nominal and acceptance currents were successfully performed. A slow ramp down was initiated after holding the current for 75 min at acceptance level due to the liquid He level dropping below the threshold. A quench was observed at 16525 A (27.36 MIIts) in the MQXFA03 magnet during the subsequent ramp, but the magnet reached the acceptance current without any training afterward.

The requirement of ramping down from nominal current with ramp rate of 150 A/s was also verified during the second cooldown. Because the issue with the liquid helium level instabilities was not yet resolved, the LQXFA/B-01 cryo-assembly was ramped again to the nominal current at 4.2 K and then we proceeded with magnetic measurements at 1.9 K. Detailed results of the magnetic measurements can be found in [7].

Several ramps to the nominal current were performed during magnetic measurements. Two trips were observed because of the still ongoing cryogenic issues. The longest run for holding current during the magnetic measurements was 75 min.

The cause of instability in the liquid helium lines was finally understood and the problem was resolved at the very end of the second thermal cycle. The five-hour holding test at the nominal current of 16230 A was successfully completed.

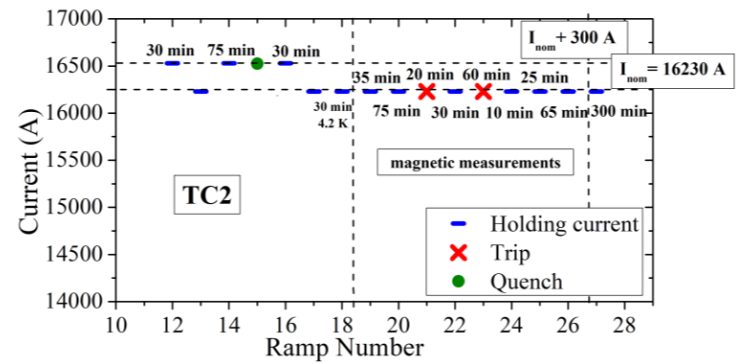

**Fig. 6.** Quench performance during the second thermal cycle (TC2). The black vertical lines mark the range of ramps performed at 4.2 K for magnetic measurements.

V. CONCLUSION

The quench performance of the first pre-series LQXFA/B-01 cryo-assembly has been successfully verified at the Fermilab's horizontal test facility. Liquid helium level instabilities and power supply issues made the test longer than expected. The cryogenic issues were identified and fixed, allowing the completion of the horizontal test.

All the quench performance requirements were met. The LQXFA/B-01 reached acceptance current without a spontaneous quench though during the test each magnet experienced a spontaneous quench. Ramp up with 30 A/s and ramp down with 150 A/s, as well as the 5-hour holding current test were successfully validated. The measured splice resistance values were below 1 n$\Omega$.